\documentclass[12pt]{article}

\usepackage{graphicx}

\def\rdots{\mathinner{\mkern1mu\raise1pt\vbox{\kern1pt\hbox{.}}\mkern2mu
   \raise4pt\hbox{.}\mkern2mu\raise7pt\hbox{.}\mkern1mu}}
\newcommand{\Z}{{\rm Z\kern-.35em Z}}
\newcommand{\bP}{{\rm I\kern-.15em P}}
\newcommand{\Q}{\kern.3em\rule{.07em}{.65em}\kern-.3em{\rm Q}}
\newcommand{\R}{{\rm I\kern-.15em R}}
\newcommand{\h}{{\rm I\kern-.15em H}}
\newcommand{\C}{\kern.3em\rule{.07em}{.65em}\kern-.3em{\rm C}}
\newcommand{\T}{{\rm T\kern-.35em T}}

\newcommand{\be}{\begin{equation}}
\newcommand{\ee}{\end{equation}}

\newcommand{\la}{\lambda}

\newcommand{\ra}{\rightarrow}

\begin{document}

\openup 1.5\jot

\centerline{Dimer $\la_d$ Expansion, Dimension Dependence of $\bar J_n$ Kernels}

\vspace{1in}
\centerline{Paul Federbush}
\centerline{Department of Mathematics}
\centerline{University of Michigan}
\centerline{Ann Arbor, MI 48109-1043}
\centerline{(pfed@umich.edu)}

\vspace{1in}

\centerline{\underline{Abstract}}

In previous papers an asymptotic expansion for the dimer $\la_d$ of the form 
\[	\la_d \sim \frac 1 2 \ ln (2d) - \frac 1 2 + \frac{c_1} d + \frac{c_2}{d^2} + \cdots \]
was developed.  Kernels $\bar J_n$ were a key ingredient in the theory.  Herein we prove $\bar J_n$ are of the form
\[	\bar J_n = \frac {C_r}{d^r} + \frac {C_{r+1}}{d^{r+1}}  \cdots + \frac {C_{n-1}}{d^{n-1}}	\]
with $r \ge n/2$.

\vfill\eject

In a series of papers a new technique was introduced and applied for extracting the asymptotic behavior of $\la_d$ in the dimer problem. [1], [2], [3].  A central ingredient in this work are the kernels $\bar J_n$ introduced in [1].  These have been computed for $n=1,...,6$ in [2].  We believe it will be exceedingly difficult to compute the $\bar J_n$ for $n>6$ (except in one dimension, $d=1$).  Once the $\bar J_n$ are known up to some order $n$, it is a simple computation (at least by computer) to extract the asymptotic behavior of $\la_d$ (up to the $1/d^{n/2}$ term if $n$ is even).  Establishing a mathematically rigorous foundation for this asymptotic theory seems a challenging task for the mathematical physicist, and will certainly involve developing good bounds on the $\bar J_n$.  Herein we study the dependence of $\bar J_n$, for a given $n$, on the dimension $d$.  We will prove that
\be	\bar J_n = \frac {C_r}{d^r} + \frac {C_{r+1}}{d^{r+1}}  \cdots + \frac {C_{n-1}}{d^{n-1}}	\ee
with $r \ge n/2$.  Equations (22) - (27) of [2] are examples of this form for the $\bar J_n$, and, so to speak, are 'tests' of this theorem.

$J_n$ is defined by equation (28) of [1] and the small modification to given $\bar J_n$ by a remark after (33) of [1].  We need put ourselves in the setting from [1].  We have from (10) of [1]
\be	v = f - f_0 \ee
where the $f's$ are functions on tiles
\begin{eqnarray}
f &=& \Bigg\{ 
\begin{array}{ll}
\frac 1{2d} & {\rm on \ dimers} \\
 \\
0 & {\rm otherwise} 
\end{array}
\\
f_0 &=& \frac{-1}{N-1} \ {\rm all\  tiles}
\end{eqnarray}
recalling tiles may be disconnected (being defined as equivalence classes of two element sets of vertices under translation).  We associate to a tile layed down covering vertices $i$ and $j$ two graphs, see Figure 1; solid edges correspond to the $f$, dashed edges to $f_0$.  Each $v$ we deal with will thus be decomposed into an $f$ and an $f_0$, a solid line and a dashed line.  $J_n$ involves a sum over $n$ tiles layed down with overlaps, so they cannot be divided into two non overlapping subsets.  Figure 2 represents a contribution to $J_4$ with one tile covering vertices $i_1$ and $i_2$, one tile covering vertices $i_3$ and $i_4$, and two tiles covering vertices $i_2$ and $i_3$.  (Different vertices in a graph are of course understood as distinct.)  There will be 15 other graphs associated to the same term in $J_4$ with some of the lines in Figure 2 replaced by dashed lines.  For purposes of this paper it is unnecessary to worry about such problems as whether the two edges joining $i_2$ and $i_3$ are distinguishable or not.

Two graphs are said to have the same {\it topology} if one can be changed to the other by changing the labelling of vertices and changing some solid lines to dashed lines and visa versa.  The {\it weight} of a graph is the product of $f^af^b_0$ where the graph has $a$ solid lines and $b$ dashed lines times the product of characteristic functions requiring the vertices bounding each solid line of the graph to be nearest neighbors.  The {\it weighted sum} of a certain set of graphs is the sum of the weights of the graphs in the set.

\bigskip
\bigskip

\noindent
\underline{Example 1}.  We consider $\cal S$ to be the set of graphs in Figure 1 with $i$ fixed, and $j$ taking all possible values (recall $j \not= i$).  Then the weighted sum of the set $\cal S$ of graphs is given by
\be \frac 1{2d} \cdot 2d - \frac 1{(N-1)} \cdot (N-1) = 0.		\ee
The first term on the left is the sum of the corresponding solid line graphs.  Here there are $2d$ choices for $j$ with non-zero contribution.  The second term the dashed line graphs.  The computation in this example corresponds to the fact that
\be	\bar J_1 = 0	\ee

\bigskip
\bigskip
\bigskip

\noindent
\underline{Fact 1}.  The $\psi'_c$ from (28) of [1] the definition of $J_n$, does not depend on the dimension $d$, but only on the topology of the corresponding graph.

\bigskip
\bigskip

This is immediate from the structure of $\psi'_c$, see equation (2.7) of [4] and the associated discussion.

\bigskip
\bigskip

There follows from Fact 1:

\bigskip
\bigskip

\noindent
\underline{Fact 2}.  To prove our theorem it is sufficient to prove that the weighted sum of all graphs with $n$ lines, of the same topology, with one vertex fixed, the others summed over, is of the form of the right side of (1).
\bigskip
\bigskip

We are interested in $\bar J_n$ in the limit $N \ra \infty$.  There follows that certain graphs need not be considered, having contributions that vanish in this limit.  In particular:

\bigskip
\bigskip

\noindent
\underline{Fact 3}.  Graphs with dashed lines inside closed loops of the graph make contributions vanishing in the  $N\ra  \infty$ limit.

\bigskip
\bigskip

\noindent
\underline{Example 2}.  Figure 3 contains some graphs that may be omitted because of Fact 3.  The computation of the contribution of the third of these graph types is easily seen to be:
\be  \left( \frac{-1}{N-1} \right)^2 \ \cdot \ (N-1) = \frac{+1}{N-1} 
\begin{array}[t]{c}
{\displaystyle\longrightarrow} \\
{N \ra \infty} 
\end{array}
 \ 0	\ee

\bigskip
\bigskip

\noindent
\underline{Fact 4}.  The weighted sum of all graphs with a certain topology and fixed number, $a$, of solid lines (one vertex fixed, others summed over) is of the form:
\be \frac 1{d^a} \ \cdot \ ({\rm Polynomial\ in} \ d).	\ee

\bigskip
\bigskip

This follows from the discussion surrounding equation (5) in [5], and the observation that the upper limit of the sum in this equation may be taken to be $n$.  I thank Gordon Slade for making this argument known to me.  Slight adaptation makes the result in [5] apply in our situation.

\bigskip
\bigskip

We turn to a proof of our theorem.  We first consider topological forms of graphs that we call {\it nondegenerate}.  A graph is nondegenerate if each of its edges lies in some closed loop.  Of the graphs in figures 1 through 4, the nondegenerate graphs are the second and third graph of Figure 3 and the graph of Figure 4.  We first note by Fact 3 that we need consider only solid line edges.  Using Fact 4 we can deduce the weighted sum of graphs of a nondegenerate topological type with $n$ edges (keeping one vertex fixed and summing over the others) has the form on the right side of (1) provided one has the following bound:

\bigskip
\bigskip

\noindent
\underline{Sum Bound}.  The weighted sum, $M$, of graphs of a nondegenerate topological type with $n$ edges (keeping one vertex fixed and summing over the others) satisfies
\be M \le \frac c{d^{n/2}}	\ee
\bigskip
\bigskip

We shall say a vertex {\it hits} a set of edges if it is in the boundary of at least one edge of the set.  We consider a graph of the nondegenerate topological type we are considering, and let vertex $i_0$ be the vertex we will fix in the sum.  We split the edges of the graph into sets
\be	{\cal S}_1, {\cal S}_2, ..., {\cal S}_t	\ee
such that
\be	\bigcup_i \ {\cal S}_i	\ee
is the set of all edges of the graph.  The sets are disjoint.
\be	{\cal S}_i \cap {\cal S}_j = \emptyset, \ \ \ i \not= j		\ee
$i_0$ hits the edges in ${\cal S}_1$.  For each $j > 1$ some vertex $i_j$ hits both the edges in ${\cal S}_j$ and the edges in
\be	\bigcup_{i < j} \ {\cal S}_i	\ee
Each ${\cal S}_j$ is either the set of edges in a closed path of the graph, or is a path whose two boundary points both hit the set in (13).  (${\cal S}_1$ must be a closed path.)

\bigskip
\bigskip

\noindent
\underline{Example 3}.  In Figure 4 we may take ${\cal S}_1$ to be union of edges $A,B,C,D$, ${\cal S}_2$ to be union of edges $E,F,G$, and ${\cal S}_3$ to be single edge $H$.

\bigskip
\bigskip

Using the decomposition of the previous paragraph, we first sum over the vertices, other than $i_0$, that hit ${\cal S}_1$; then the vertices that hit ${\cal S}_2$, but not ${\cal S}_1$.  At the $jth$ stage, the vertices hitting ${\cal S}_j$ but not the set of (13).  We now use ideas parallel to those of Fact 4.  Let there be $\ell_j$ edges in ${\cal S}_j$.  As one moves along the path of the edges of ${\cal S}_j$, summing over the $(\ell_j - 1)$ new vertices, and determining the edges along the path, each time an edge lies in an entirely ''new' direction, another edge of the same path must be in minus that direction.  Thus the number of terms in the sum over vertices in the path is bounded by
\be	cd^{\ell_j/2}	\ee
This proves the Sum Bound, equation (9).

We now note that referring to Fact 2, we need only show one has the same bound, equation (9), for degenerate graphs to have completed the proof of our theorem.  We call edges, not in a closed loop, {\it free edges}.  We let a wavy line in a figure denote a free edge, either a solid line or dashed line.  For a given free edge, we define its {\it ground} to be the edges and vertices of the graph connected to the fixed $i_0$ by paths not including the given free edge.  And its {\it float}, to be edges and vertices of the graph not connected to $i_0$ by paths not including the given free edge.  In Figure 5 we have illustrated schematically the ground and float of the free edge, the wavy line.  In the first diagram of Figure 6, the float consists of a single vertex, the ground, two vertices and an edge.

We will show that the weighted sum of a degenerate graph can be replaced by a linear combination of weighted sums of nondegenerate graphs.  We show this by induction on the number of free edges.  The $n=1$ case and the induction step are the same, each reducing the number of free edges that need to be considered.

We consider the graph of Figure 5.  The first sum we perform will be over the position of vertex $j$.  In performing this sum the vertices in the ground will be kept fixed, but the vertices in the float all moved in sync with $j$.  That is, we have changed variables, vertex $k$ in the float will be described by a variable $(k-j)$, that is kept fixed as $j$ is summed over.

\bigskip
\bigskip

\noindent
\underline{Important Observation}.  If the float never bumps into the ground as $j$ is summed over the sum is zero.

\bigskip
\bigskip

The summation done before for Figure 1 was a trivial case of this observation.  When the wavy line represents a dashed line it doesn't matter if the float bumps into the ground, such cases give no input in the $N \ra \infty$ limit.  Using our observation, we may replace the sum over the vertex $j$ in our diagram by working with graphs where the wavy line is replaced by a solid line and taking minus the sum over all ways the vertices of the float may match with vertices of the ground (keeping distinct vertices distinct).  Thus the sum over $j$ in the first diagram of Figure 6 may be replaced by minus the second diagram of Figure 6.  Remaining sums may then be performed.

\bigskip
\bigskip

\noindent
\underline{Final Note}.  Many graphs one may deal with make zero contribution in any weighted sum.  Thus with the first graph of Figure 6, one could have matched vertex $j$ of the float to vertex $i$ of the ground, getting a graph we did not draw because it has zero weight.

\vfill\eject

\centerline{\underline{References}}

\begin{itemize}
\item[[1]] Paul Federbush, ``Hidden Structure in Tilings, Conjectured Asymptotic Expansion for $\la_d$ in Multidimensional Dimer Problem", \ arXiv : 0711.1092V9 [math-ph].
\item[[2]] P. Federbush, ``Dimer $\la_d$ Expansion Computer Computations", arXiv: 0804.4220V1 [math-ph].
\item[[3]] P. Federbush, ``Dimer $\la_3 = .453\pm.001$ and Some Other Very Intelligent Guesses", arXiv:0805.1195V1 [math-ph].
\item[[4]]  David C. Brydges, ``A Short Course in Cluster Expansions, Phenomenes critiques, systems aleatoires, theories de gauge, Part I, II" (Les Houches, 1984), 129-183, North Holland, Amsterdam, 1986.
\item[[5]] Nathan, Clisby, Richard Liang, Gordon Slade, ``Self-avoiding walk enumeration via the lace expansion", {\it J. Phys. A : Math. Theor.} 40 (2007) 10973-11017.

\end{itemize}

\newpage
\begin{center}
\includegraphics[scale=0.7]{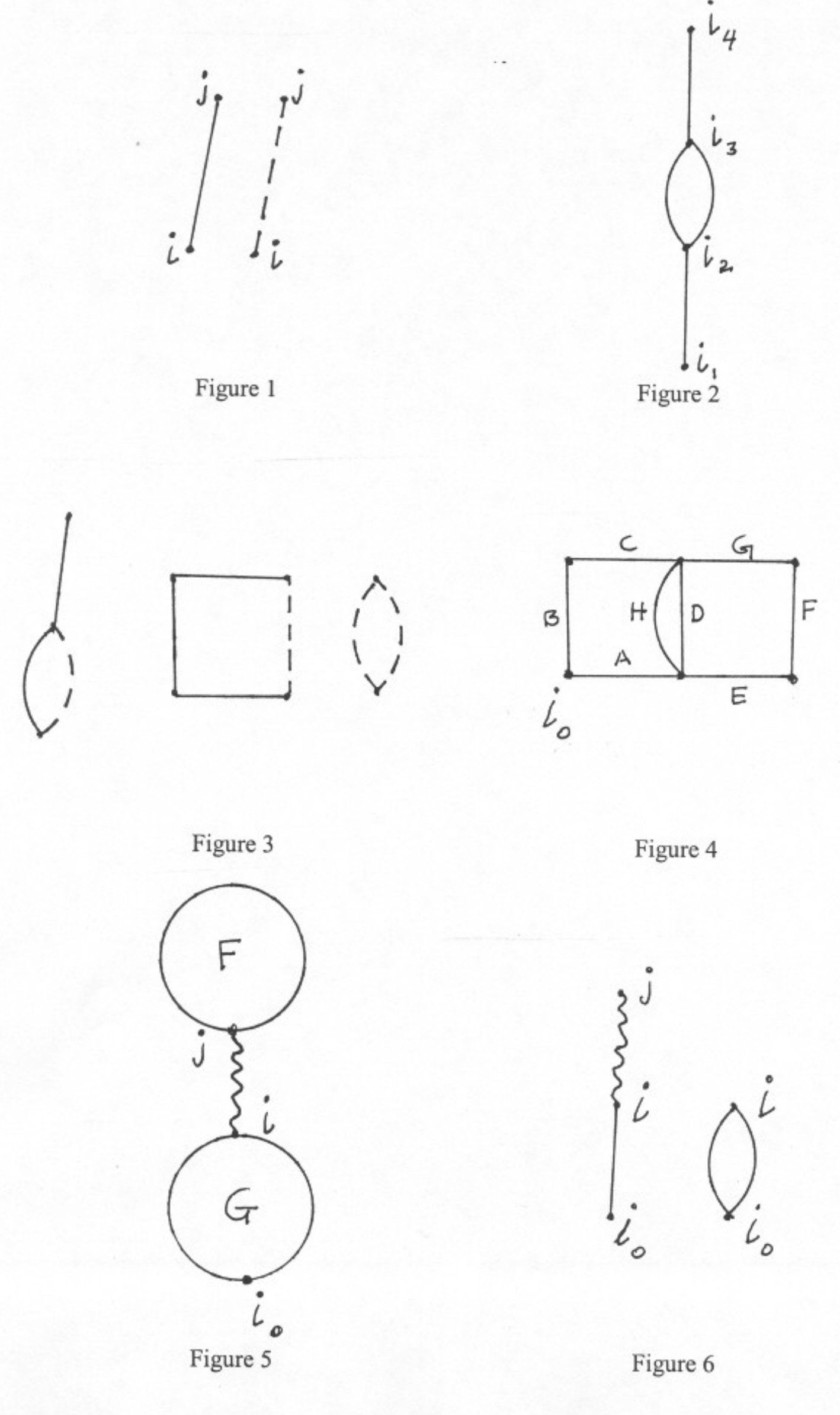}
\end{center}

\end{document}